\documentclass[%
reprint,
superscriptaddress,
showpacs,
nofootinbib, nobibnotes, 
 amsmath,
 amssymb,
 aps,
 pre
]{revtex4-1}

\usepackage{times}

\usepackage{microtype}
\usepackage{blindtext}
\usepackage{dcolumn}
\usepackage{bm}
\usepackage{color, soul, colortbl} 
\usepackage[colorlinks=true,
						linkcolor=blue,
						urlcolor=blue,
						citecolor=blue,
						bookmarks=true,
						pdfborder={0 0 0}]{hyperref}

\usepackage{xcolor}
\usepackage{mdframed} 

\usepackage{graphicx,epsfig}
\usepackage{subfigure}
\usepackage{amsmath,amsfonts}
\usepackage{xfrac}
\usepackage{enumerate}
\usepackage[overload]{empheq}
\usepackage{mathtools}
\usepackage{cases} 
\usepackage{cancel} 

\begin{document}
\title{On the dual nature of adoption processes in complex networks}

\author{Iacopo Iacopini}
	\email[E-mail: ]{iacopiniiacopo@gmail.com}
    \affiliation{School of Mathematical Sciences, Queen Mary University of London, London E1 4NS, United Kingdom}
    \affiliation{Centre for Advanced Spatial Analysis, University College London, London W1T 4TJ, United Kingdom}
    \affiliation{Aix Marseille Univ, Universit\'e de Toulon, CNRS, CPT, Marseille, France}
\author{Vito Latora}
	\affiliation{School of Mathematical Sciences, Queen Mary University of London, London E1 4NS, United Kingdom}
	\affiliation{The Alan Turing Institute, The British Library, London NW1 2DB, United Kingdom}
	\affiliation{Dipartimento di Fisica ed Astronomia, Universit\`a di Catania and INFN, I-95123 Catania, Italy}
	\affiliation{Complexity Science Hub Vienna, A-1080 Vienna, Austria}


\begin{abstract} 
Adoption processes in socio-technological systems have been widely studied both empirically and theoretically. The way in which social norms, behaviors, and even items such as books, music or other commercial or technological product spread in a population is usually modeled as a {\em process of social contagion} in which the agents of a social system can infect their neighbors on the underlying network of social contacts. More recently, various models have also been proposed to reproduce the typical dynamics of a {\em process of discovery}, in which an agent explores a space of relations between ideas or items in search for novelties. In both types of processes, the structure of the underlying networks, respectively the network of social contacts in the first case, and the network of relations among items in the second one, plays a fundamental role. However, the two processes have been traditionally seen and studied independently. Here, we provide a brief overview of the existing models of social spreading and exploration and of the latest advancements in both directions. We propose to look at them as two complementary aspects of the same adoption process: on one hand, there are items spreading over a social network of individuals influencing each other, on the other hand, individuals explore a network of similarities among items to adopt. The two-fold nature of the approach proposed opens up new stimulating challenges for the scientific community of network and data scientists. We conclude by outlining some possible directions that we believe may be relevant to explore in the coming years.
\end{abstract}
\maketitle

\section{Introduction}

Networks constitute the backbone of complex systems, from the human brain to computer communications, from metabolic and protein systems to online and offline social systems. Characterizing their structure improves our ability to understand the physical, biological and social phenomena that shape our world~\cite{newman2010networks,barabasi2016network, latora2017complex,battiston2020networks}. The structure of the network plays in fact a major role in the dynamics of a complex system and characterizes both the emergence and the properties of its collective behaviors~\cite{barrat2008dynamical, vespignani2012modelling}.  In particular, over the last twenty years, networks have been extensively used to model human behavior, and such studies have attracted the attention of sociologists, economists, physicists, and computer scientists. The network approach, eventually integrated with the opportunities offered by the newly available data sources~\cite{salganik2019bit, ledford2020facebook}, has largely contributed to the growth of new interdisciplinary fields such as those of sociophysics~\cite{castellano2009statistical,sen2014sociophysics, baronchelli2018emergence} and computational social science~\cite{Lazer721, conte2012manifesto, golder2014digital,lazer2020computational}.

In this paper, we focus on network approaches that aim at capturing and modeling the fundamental mechanisms behind the {\em social dynamics of adoption}~\cite{granovetter1977strength, valente1995network}. The processes through which humans discover and adopt novel items--where by items we indicate not only artefacts or new technological or commercial products, but also concepts, ideas, social norms and behaviors--can be described in two radically different ways, both involving the presence of a complex network, whose nature is different in the two cases.

Indeed, on one hand item adoption can be seen as a {\em contagion dynamics} over a social network of individuals influencing each other through their social connections. On the other hand it can be described as an {\em exploration dynamics} over a network of similarities among the different possible items that an individual can adopt.\\
In the first case a single item (a single product, idea, or behavior) is considered at once, and the transmission from one individual to another over a social system is modeled as an epidemic-like spreading process ~\cite{goffman1964generalization, pastor2015epidemic} across the links of a social network~\cite{bass1969new, centola2007complex, centola2010spread}. Hence, the focus here is on the complex structure of the underlying social network.
In the second case, the main focus is instead on the network of existing relations between different items~\cite{toole2015coupling}. Hence, the modeling attention is shifted towards the cognitive processes through which single individuals explore the space of different possibilities and produce sequences of explored items in search of novelties~\cite{bendetowicz2018two, siew2019cognitive,zhou2020growth}.  In this latter way of interpreting adoption dynamics, different exploration (and innovation) models have been proposed to replicate the process of exploration according to which one idea, concept or item leads to another, and a discovery can trigger further ones~\cite{tria2014dynamics, iacopini2018network}.
Here, we present a brief overview of two different types of mechanisms that can contribute to a process of adoption in socio-technological systems, namely social contagion on the one hand, and item exploration on the other. More specifically, in Sec.~\ref{sec:contagion} and Sec.~\ref{sec:discovery} we will discuss the two different aspects of adoption dynamics, briefly describing the various modeling approaches that have been proposed and the latest advancements in both directions. In Sec.~\ref{sec:coupling} we will elaborate more on this duality, illustrating the potential of constructing more elaborate models that consider at the same time the two types of processes. Finally, in Sec.~\ref{sec:conclusions} we will highlight various directions in which models could be expanded by implementing more elaborated mechanisms both at the level of the discovery dynamics and at that of the social contagion process.

\begin{figure*}[t]
	\begin{center}
		\includegraphics[width=0.8\textwidth]{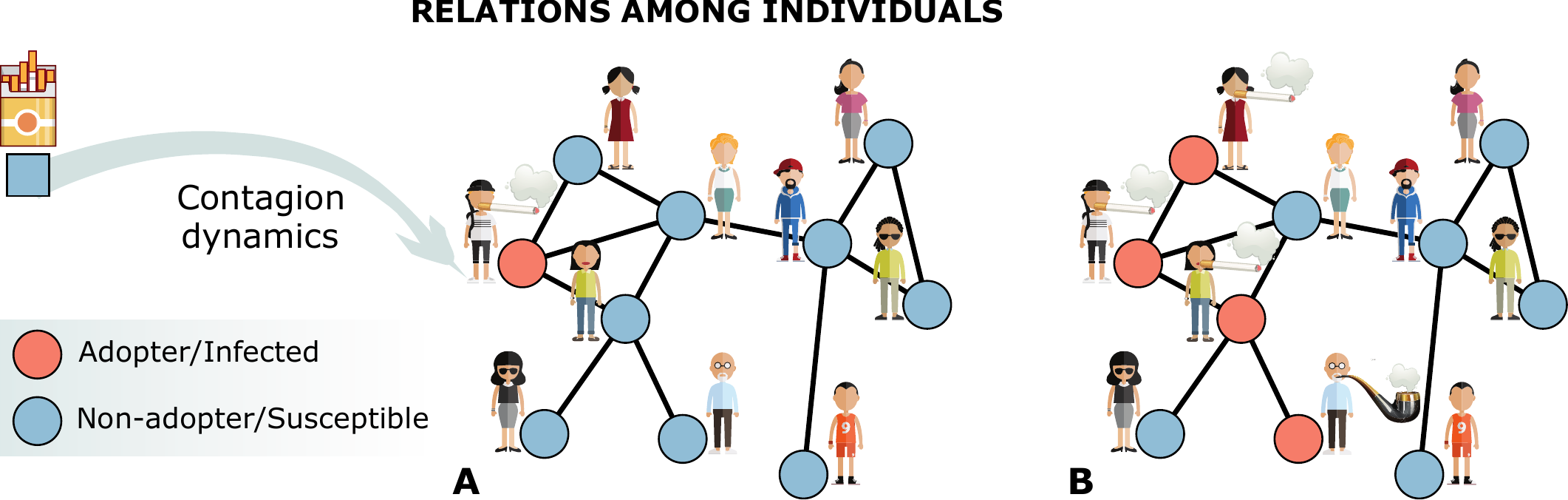}
	\end{center}
	\caption{{\bf Illustration of a contagion process.} The adoption of norms, behaviors, ideas, technological items, etc., is typically modeled as a spreading process over a network of social contacts. Red and blue nodes of the social network denote respectively the adopters (or infected individuals) and the non-adopters (or susceptible) of the item that is spreading. For example, in ({\bf A}) a smoker transmits the\textemdash bad\textemdash habit to its neighboring agents, which in turn can transmit it again through their social links ({\bf B}).}\label{fig:contagion}
\end{figure*}

\section{Modeling social contagion}\label{sec:contagion}

Quantitative studies of contagion phenomena have helped shading light on the similar dynamics at which information, viruses, knowledge, rumors, and innovations spread in a population~\cite{goffman1964generalization, daley1964epidemics, lloyd2001viruses, young2009innovation}. Contagion processes are usually mediated by interactions, that should therefore be taken into account into the modeling framework \cite{watts2002simple, de2018fundamentals, christakis2013social, gleeson2013binary}. As a consequence, as for other landmark dynamical processes widely studied within the complex systems community, the interplay between the social structure and the contagion dynamics that unfolds upon it has been the focus of many studies~\cite{christakis2013social, gleeson2016effects, pond2020complex}.\\
The mechanisms of a basic contagion dynamics are illustrated in Figure~\ref{fig:contagion}. Given a single behavior, like smoking~\cite{christakis2008collective}, the diffusion of the habit can be thought as a spreading process over a network of social relationships in which individuals influence each other towards the adoption of the behavior, thus going from Figure~\ref{fig:contagion}A to Figure~\ref{fig:contagion}B. In this basic representation, the spreading dynamics through a population of interacting individuals is similar to the one of pathogens~\cite{goffman1964generalization}, and the typical modeling approach, akin to the one of disease spreading, relies on compartments into which individuals are classified according to their status: adopters/infected (I) and non-adopters/susceptible (S). Susceptible-Infected (SI) is thus the simplest model in this setting, where an individual can move--after a contact with an I--from compartment S to I with a given probability of infection. In the Susceptible-Infected-Susceptible (SIS) variation, the infection process is reversible, and infected individuals can recover. This transition involves a second parameter, the recovery rate.
These models can be very informative, but their simple mechanisms work only at a first approximation. In fact, when dealing with human systems, there is a variety of behavioral aspects influencing the social dynamics that cannot be overlooked~\cite{hodas2014simple}. In many cases the dynamics cannot be simply explained in terms of basic disease epidemics models, which would result too reductive. Instead, the social nature of the contacts mediating these processes deserves special attention, calling for ad-hoc modeling adjustments and tailored experimental techniques to measure social effects~\cite{aral2009distinguishing, aral2011creating}.

Along this line, recent investigations have empirically shown that {\it simple contagion} rules (SI$\longrightarrow$2I) are not appropriate to describe the more complex mechanisms of social influence that are at work when humans interact~\cite{christakis2008collective, centola2010spread, onnela2010spontaneous, monsted2017evidence}. This evidence, mostly provided by digital traces, relates to different contexts, ranging from for the adoption of applications~\cite{onnela2010spontaneous, karsai2014complex} and technologies~\cite{bandiera2006social, oster2012determinants} to the spreading of obesity~\cite{christakis2007spread} and happiness~\cite{fowler2008dynamic}.\\
These considerations gave rise to new streams of research focused on translating theories coming from the social sciences into mechanistic models. Among these, {\it complex contagion} is a particularly popular theory according to which--differently from simple contagion--multiple stimuli are necessary to trigger a behavioral change in a population of interacting individuals~\cite{centola2010spread, centola2018behavior}. Current efforts in this direction have been summarized by Guilbeault et al. in Ref.~\cite{guilbeault2018complex}. Although supported by a mounting body of empirical studies~\cite{centola2010spread, ugander2012structural, karsai2014complex, monsted2017evidence, lehmann2018complex}, complex contagion is not the only theory out there, and alternative mechanisms have been theorized. For example, \citet{ugander2012structural} proposed {\it structural diversity}, a local measure of the neighborhood of a node, quantified in terms of number of connected components having at least one adopter. When empirically tested on data of adoption of online platforms upon invitation, this measure turned out to be a better predictor of the probability of adoption with respect to more conventional measures like the number of adopters among the peers. Complex contagion and structural diversity have also been tested against {\it embeddedness} and {\it tie strength} theories, in which friendship overlap and intensity are the key drivers of social contagion instead~\cite{aral2014tie, aral2017exercise}.

While most of the works mentioned so far focused on the adoption mechanism from the single-sided transition that leads to the adoption, some attention has been also given to the opposite process, in which adopters abandon the new product or technology and become ``susceptible" again (I$\longrightarrow$S). Recently, it has been shown that differences between the recovery rates of the nodes, i.e., considering heterogeneous distributions of parameters instead of constant, can also dramatically change the epidemiological dynamics~\cite{de2020impact, darbon2019disease, brett2019spreading}. In addition, models of {\it complex recovery}, in which the social influence mechanism acts on the recovery rule rather than on the infection one, showed that this change of perspective might lead to explosive adoption dynamics~\cite{iacopini2020multilayer}. This behavior is especially pronounced in spatial systems, whose effects on the contagion dynamics have also been the focus of several studies~\cite{strang1993spatial, toole2012modeling, lengyel2020role, davis2020phase}.\\
Yet, in human communication interactions can occur in groups of three or more agents, and often cannot be simply factored into a collection of dyadic contacts. Expanding the pairwise representation given by graphs in favor of a non-pairwise one, like simplicial complexes or hypergraphs, is a recent research direction that finds in social systems a particularly suitable playground~\cite{battiston2020networks}. A paradigmatic example is the model of {\it simplicial contagion}, that shows how the inclusion of these higher-order group interactions can dramatically alter the spreading dynamics and lead to the emergence of novel phenomena, such as discontinuous transitions and bi-stability~\cite{iacopini2019simplicial}. Similar results can be also found when hypergraphs are used to encode social patterns underneath the spreading process instead~\cite{de2020social, landry2020effect}. 

\section{Modeling discovery processes}\label{sec:discovery}
Novelties are part of our daily life. The discovery dynamics at which an individual consumes goods or listens to songs can be described, using the words by Thomas Kuhn~\cite{kuhn1979essential}, in terms of the {\em essential tension} between exploitation and exploration. This eternal trade-off recurs in a variety of different systems. For example, people move between different locations, mostly switching between already known places, but also visiting new ones from time to time~\cite{pappalardo2015returners, yan2017universal, alessandretti2018understanding, alessandretti2018evidence}. The individual propensity towards ``uncharted seas" enters in each discovery processes and enables classifications, such as the returners versus explorer dichotomy for human mobility~\cite{pappalardo2015returners}. If we think of each visit of a place, listening of a song, or, more in general, collection of an item as the addition of a symbol to a symbolic sequence, the series of actions of an individual (agent) can be represented as a sequence that grows in time, over an alphabet that represents a space of possibilities. Symbolic sequences have a long history in text analysis, but recently sequences of item adoptions have been used to study human behavior, leveraging sequences of purchases as tracked by credit card data~\cite{di2018sequences} or supermarket fidelity cards~\cite{aiello2019large, aiello2020tesco}. Any process involving individuals and objects that can be encoded into a sequence of actions can be framed in this way. People explore a space, adopt new items, and often return on their steps. Every time a new item enters the sequence it can thus be seen as a novelty.\\
\begin{figure*}[t]
	\begin{center}
		\includegraphics[width=0.8\textwidth]{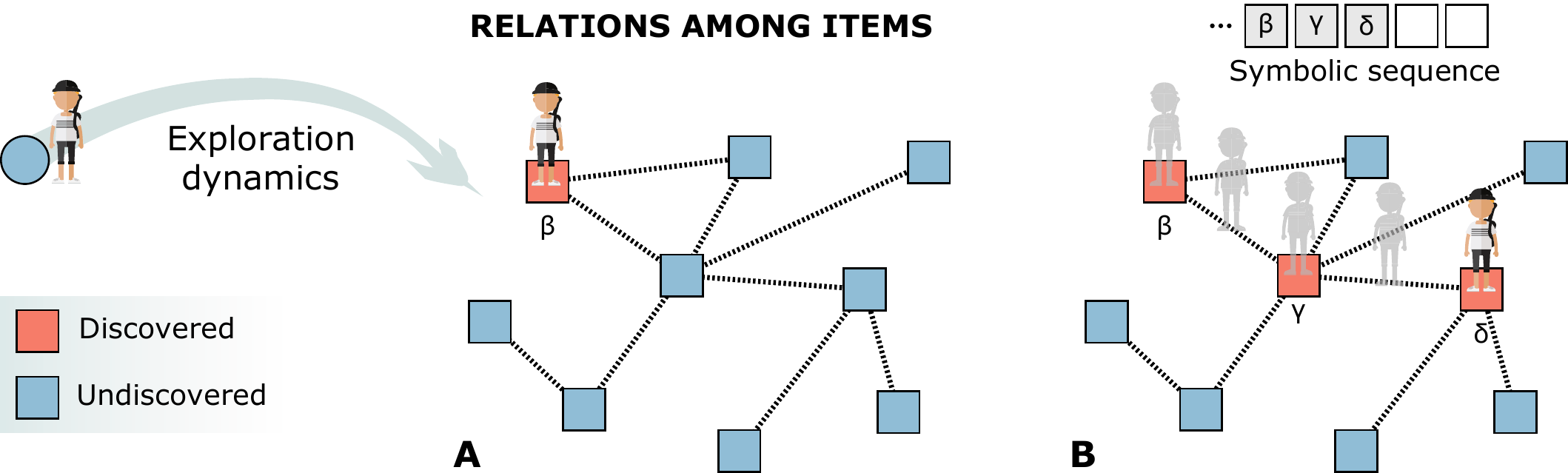}
	\end{center}
	\caption{ {\bf Illustration of an exploration process.}
		The cognitive process through which an individual agent explores the space of
		possibilities in search of novel items (novel ideas, technological discoveries) is usually modeled as a walk over a network of relations similarities or proximity) among items. For example, in ({\bf A}) the agent discovers item $\beta$ and then continues the exploration over the links of the network by sequentially moving to node $\gamma$ and then to node $\delta$. In ({\bf B}) three items have been discovered, and the exploratory walk can be seen as a sequence of symbols representing the visited nodes.}\label{fig:exploration}
\end{figure*}
This precise mechanism of exploration and exploitation becomes particularly relevant at the collective level, where novelties can be interpreted as innovations~\cite{johnson2010good}. In fact, the first discovery by any individual of a population represents a novelty for everybody. In this scenario, the essential tension between tradition and innovation has been the focus of many studies that analyzed the collective action of researchers determining scientific progress~\cite{foster2015tradition, clauset2017data, murdock2017exploration, aleta2019explore, hofstra2020diversity}. On the same line, patent data have been largely used to explore the dynamics of technological ecosystems~\cite{coccia2014driving, pichler2020technological} with the aim of predicting the innovation dynamics and eventually detecting the best strategies that could influence the rate of innovation~\cite{fink2017serendipity, coccia2014driving, Finkeaat6107}.\\
Researchers have been tackling the problem of the emergence of innovation from different angles. For example, some studies have been focusing on the dynamics of substitutive systems, in which the new always replaces the old~\cite{jin2019emergence}. Here instead, we keep the focus on the dynamics leading to the emergence of the new; we frame the problem in a cumulative way, such that the new, intentionally very broadly defined, always comes as an addition to the existing. More precisely, the existing environment is actually a necessary condition that paves the way to the emergence of the new. In fact, from Parmenides to modern evolutionary biology, ``nothing comes from nothing" is a dictum at the essence of each process involving real-world systems. Thus, even if we neglect that new items might arise from the re-combination of existing ones~\cite{thurner2010schumpeterian, youn2015invention}, there is still an essential ingredient that models should take into account, that is the structure underneath these items which determines the way in which individuals can navigate it~\cite{correa2019semantic, lynn2020human} moving from one item to the next. For example, knowing the bestseller of a book writer is often a condition that puts us in the position of deepening our research towards minor novels of the same author. Similarly, in scientific production, early research on the properties of random walks enabled further studies on biased random walks.
In this setting, one can think of knowledge as an unexplored space of relationships between concepts and objects to be discovered by\textemdash more or less ``innovative"\textemdash investigations and experiments~\cite{rzhetsky2015choosing, zurn2020network}. These could be interpreted either as an exploration processes of an abstract space of concepts, ideas, items, etc.~\cite{cattuto2009collective}, and as a knowledge acquisition process~\cite{rodi2015optimal, de2017knowledge, lima2018dynamics}, like people acquiring information through online searches~\cite{rodi2017search, lydon2020hunters}. Research experiments in this direction aim at understanding how humans explore and build mental representations of these spaces through the experience of sequential items. Ref.~\cite{lynn2020humans} is a review of the recent efforts, framed under the graph learning paradigm. An important aspect in this scenario is that the structure of this space does matter, since some portions of the space are only visible from certain positions. This concept resonates with the evolutionary theory of the {\it adjacent possible} developed by Stuart Kauffman~\cite{kauffman1996investigations}. According to this framework, we can split the knowledge space into what has already been discovered (the actual) and what is left to explore (the possible). Notice, however, that only one tiny fraction of the possible is achievable from the actual, and this is precisely the adjacent possible, that is situated one step away from what is already known.

Many models have been put forward. At their essence, there is often a reinforcement mechanism, akin to the {\it rich-get-richer} paradigm~\cite{perc2014matthew}, that accounts for self-reinforcing properties. This is an essential ingredient that allows to recover the scaling laws emerging from discovery processes in real-world systems~\cite{tria2014dynamics, monechi2017waves}, like the well-known Heaps', Zipf's, and Taylor's laws~\cite{heaps1978information, zipf2016human, Tria_2018}, for example via sample-space-reducing mechanisms~\cite{mazzolini2018heaps}. The Yule process~\cite{yule1925ii} is one of the first mechanisms employed to generate the empirically observed power laws. From there, many processes with reinforcement have been developed~\cite{pemantle2007survey}. At their root, there are the famous standard urn processes~{\cite{johnson1977urn}, like the ones of P{\'o}lya and Hoppe~\cite{polya1930quelques, hoppe1984polya}. However, these basic processes have been slowly modified and tuned with empirical data in order to better capture the observed patterns. An example is the generalization of the Yule-Simon process developed to mimic the dynamics of collaborative tagging, where online users associate tags (descriptive keywords) to items, generating fat-tailed frequency distributions of tags~\cite{cattuto2007semiotic}.\\
Later, the urn framework has been further enriched in order to account for the dynamics of correlated novelties. In fact, empirical traces of human activities show that discoveries come in clusters, and the symbolic sequences generated by discovery processes are thus correlated~\cite{tria2014dynamics, monechi2017waves}. Models can mimic this behavior by letting the space grow together with the process, such that novelties increase the number of possible discoveries via triggering mechanisms. A review of these models of expanding spaces can be found in Ref.~\cite{loreto2016dynamics}. Leveraging the concept of the adjacent possible, triggering mechanisms showed good agreement with empirical data in reproducing both the scaling laws associated to the discovery processes and the correlated nature of the sequences they produce. This is the case of the urn model with triggering~\cite{tria2014dynamics} that incorporates the adjacent possible within the urn process, or the edge-reinforced random walk model~\cite{iacopini2018network} that encodes it into the topology of a network of concepts and ideas.
It is easy to see how the network representation of the space of items naturally accounts for the adjacent possible, since paths are restricted to existing connections, and the discovery (visit) of a given node can provide access to a different set of nodes not directly accessible before. This is illustrated in Figure \ref{fig:exploration}. The use of networks as an underlying structure for search strategies and navigation is strictly linked to the literature in random walks and optimal foraging~\cite{masuda2017random}, but lately it has been applied to various contexts. For example, in cognitive sciences, networks have been used at length to encode the patterns behind mental representations~\cite{thagard2005mind, borge2010semantic, baronchelli2013networks, castro2020contributions}. There, as for contagions, understanding the influence of these structures on the process of discovery that unfolds on top remains a fascinating problem.
	
\section{Coupling the dynamics of discovery and contagion}\label{sec:coupling}

\begin{figure*}[t]
	\begin{center}
		\includegraphics[width=\textwidth]{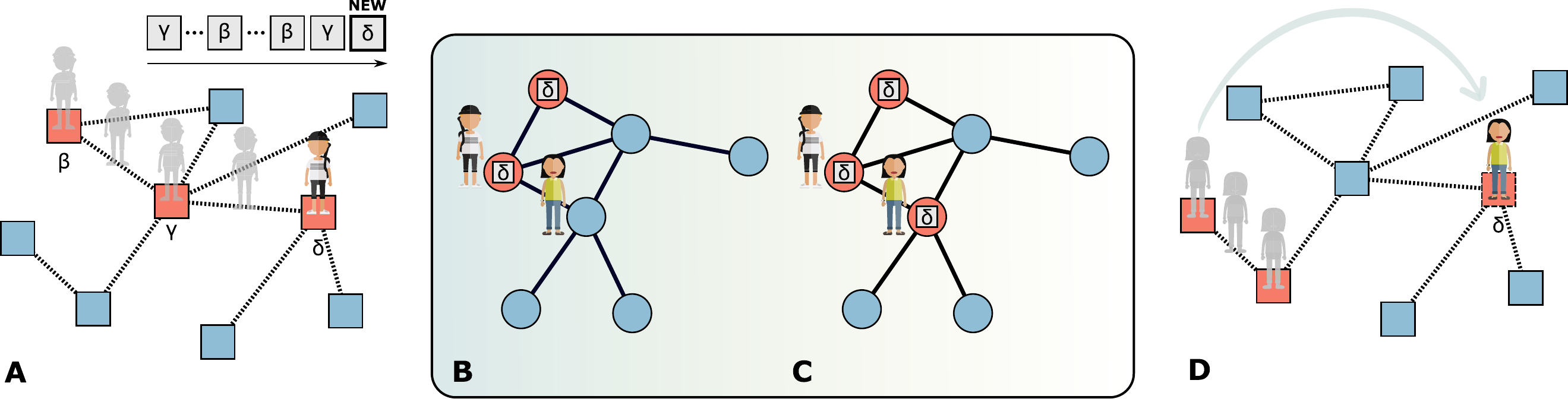}
	\end{center}
	\caption{ {\bf Coupling the dynamics of discovery and contagion.} ({\bf A}) A walker explores the space of items. Every time a node is visited the corresponding item is added to the sequence associated to the agent. Walking on $\delta$ represents the discovery of a novelty, since it never appeared in the sequence of before. ({\bf B}) At this point the new item $\delta$ can spread from the walker to his neighbors through the link of a social network following a process of social contagion. In particular, in ({\bf C}) a neighbor discovers $\delta$ through the social contagion dynamics. As a consequence, her position on the network of items is updated by means of a flight ({\bf D}).}\label{fig:themodel}
\end{figure*}
	
The two previous sections have shown that adoption processes can be approached from two different angles. However, this has turned into a sort of {\em dual nature of adoption processes}, as most of the studies have proposed either to model how an individual explores a network of items, or how one item spreads over a social network. Focusing either on one item or on one individual at a time means neglecting a fundamental aspect of socio-technological systems, their multi-agent nature.
In a scenario with non-substitutive items, the interactions between multiple items (or behaviors) simultaneously spreading on a social network can ultimately affect the process, eventually favoring or inhibiting the adoption, as it happens for the spreading of multiple infectious diseases over a population~\cite{wang2019coevolution}. This mechanism would be less pronounced in competitive environments, when at each time nodes can take only one of many possible states~\cite{fennell2019multistate}, and different states would correspond, for example, to the items agents are presently trying to share. However, even in this case, the state of the neighboring nodes could influence the adoption process as well.  In fact, people do not live and work in isolation, and social ties can shape their behaviors~\cite{mcpherson2001birds}. As such, discovery processes of different individuals are surely not independent from one another. For example, a higher similarity has been found in the mobility patterns (extracted from the sequences of visited cell phone towers) of individuals that share a social tie--when compared to the case of strangers~\cite{toole2015coupling}. In music listening habits, it has been found that going to a concert increases the amount of post-event plays of that artist for both attendees and  attendees' friends who did not even attended the concert~\cite{ternovski2020social}. Thus, these purely social effects would have also an impact on the dynamics of novelties, as it has been recently shown with a model of interacting discovery processes~\cite{iacopini2020interacting}. When urn processes (representing individual explorers) are coupled through the links of a social network, so that explorers can exploit opportunities (possible discoveries) coming from their social contacts, the position in the network affects the ability to discover novelties: individuals with large values of centrality over the social network have in general a higher discovery pace.

Despite some preliminary and promising results in this direction, a comprehensive framework that bridges together the dual nature of adoption processes, namely discovery and contagion, is still missing. 
Ideally, a general model should consider and couple together two different networks, namely the network of social relations among the agents of a social system, and the network of relations among items. As discussed in Sec.~\ref{sec:contagion}, contagion occurs through social interactions, and thus the process of contagion would use as a substrate a social network. In parallel, there could be an exploration processes with reinforcement--akin to those introduced in Sec.~\ref{sec:discovery}--that takes place over a different network of relations/similarities between items. Individuals could perform independent searches over the space of items~\cite{loreto2016dynamics, zurn2020network}, modeled for example as random walks with reinforcement. In this case, the strength of the links could vary across different explorers according to their personal history. By using edge-reinforced random walks~\cite{iacopini2018network}, walkers would share the same structure of the network of items, while the strengths of the connections would co-evolve with the dynamics of the walkers, highlighting the fact that different walkers can prefer to move towards different items (i.e. can prefer certain cognitive associations more than others) also according to their memory.
The reinforcement mechanism, which quantifies the tendency of each explorer to go back on her/his steps, reflects the fact that some individuals might be more inclined to the exploration, while others might be less keen to adventure and would rather prefer to exploit the already known~\cite{pappalardo2015returners}. Thus, while the results in Ref.~\cite{iacopini2020interacting} stress the impact of the social network, as given by the node centrality, in determining the speed at which {\it identical} individual discover new items, the impact of having heterogeneous individual ``memory" remains a fascinating aspect to be explored. In addition, due to the reinforcement, discovering new ideas would become more and more difficult, and walkers will often return on their steps. This implies that the last novelty found will remain unchanged for some time. As for processes of individual and collective attention in social and substitutive systems~\cite{jin2019emergence}, individuals will focus on a single novelty at the time. A simple way to couple discovery and contagion is illustrated in Fig.~\ref{fig:themodel}, where each individual tries to spread this last novelty to her/his neighbors through the links of the social network by means of a simple SI contagion process. If such a novelty will spreads enough, it will eventually become popular. This could be linked to recent works that have shown how simple mathematical models can accurately describe processes of topics and memes that compete for collective attention, displaying bursts and decays~\cite{weng2012competition, gleeson2014competition,lorenz2019accelerating, candia2019universal}. In a novelty-driven scenario, every time that a walker finds a novelty, she would consequently try to spread it to a neighbor in the social network. If the neighbor will adopt such a novelty (this might depend on an individual ``susceptibility''), then she would immediately move to the correspondent node on the network of items, ultimately adding long jumps to the random walk process. Without these jumps, the exploration probability of the individuals would decrease over time due to the reinforcement. However, this would be in contrast with the long-term evolution of the strains of activities that has been empirically observed on human mobility traces~\cite{alessandretti2018evidence}. Thus, while standard mechanisms to overcome this issue involve the use of finite memory and recency~\cite{barbosa2015effect}, the social benefits induced by the coupling could represent an alternative mechanism, where individuals ``stuck" with a certain set of established items may evolve their preferences due to social influence.

\section{Perspective and future directions}\label{sec:conclusions}
	
Discovery and contagion are two important processes in social systems, whose basic mechanisms and dynamical behaviors have been extensively studied both empirically and theoretically. However, with very few exceptions, these two processes have always been seen and studied separately.
	
In this perspective, we have shown that the two processes can be considered as two different aspects of a single process of adoption in a social system. The dynamics of adoption of items, which can either be concrete objects or behaviors and social norms, can indeed be seen from two different angles as: 1) a contagion process occurring over a network of individuals influencing one another, or 2) as an exploration dynamics over a network of items that an individual can adopt. This means considering respectively, either the point of view of a single item spreading over a social network of individuals, or the point of view a single individual sequentially moving from adopting one item to adopting a similar one. In this work, we have discussed such a duality, the latest modeling advancements, and we have proposed an example of how to couple together these two complementary processes of discovery and contagion in a single social adoption model.
Merging together discovery and contagion models would enable to explore the effects of social influence on the mechanisms behind the exploration of new items~\cite{iacopini2020interacting}. However, while the coupled dynamics of discovery and contagion proposed here relies on {\it simple contagion} mechanisms, future research could make use of more sophisticated contagion mechanisms, such as threshold models. It would also be interesting to investigate whether the jumps induced by the social contagion can disrupt the composition of the set of exploited items, as it happens in social networks~\cite{saramaki2014persistence}.

We hope that the ideas introduced here can represent only the first step of what what can be a long journey. In fact, different important aspects need to be further explored on both sides, and many existing models of discovery and contagion could be coupled together.\\
Regarding discovery processes, an essential mechanism that should be included in future works is the dynamics of items re-combinations. In processes of semantic associations the combination of remote ideas can lead to Eureka moments ~\cite{bendetowicz2017brain}. Similarly, in real-world systems, new items can be generated and thus discovered by combining existing ones. An example is the model of Schumpeterian economic dynamics proposed by~\citet{thurner2010schumpeterian} that relies on creation and destruction tables. More interestingly, the action of creating compounds of items could be influenced by complementary skills that become available once exploiting the social network. Some early works in this direction involve the combination of medicinal plants to generate novel drugs in an innovation model of hunter-gatherers~\cite{migliano2020hunter}, and the creation of scientific knowledge from the combination of different expertise to generate high-impact science~\cite{uzzi2013atypical}.\\
In addition, some recent advances in the foundation of complex systems have started to consider the higher-order nature of interactions in real-world systems~\cite{battiston2020networks}. While the role of these higher-order structures with respect to the dynamics of spreading and diffusion has started to be explored~\cite{iacopini2019simplicial}, most of the literature on discovery processes is still limited to pairwise representations. Instead, the fundamental units of scientific productivity are research groups and teams~\cite{guimera2005team, bonaventura2020predicting}, and thus higher-order approaches could naturally be used when studying processes of collective scientific discoveries and teamwork~\cite{torrisi2019creative, monechi2019efficient, almaatouq2020social}. In fact, scientific collaborations are the perfect example of higher-order system whose representation demands for non-pairwise building blocks~\cite{yang2015knowledge, patania2017shape}.\\
Finally, more sophisticated discovery mechanisms could be included in the models. For example, explorers could interact in different ways with each other and with the different items~\cite{cencetti2018reactive, skardal2019dynamics}. Alternatively, the exploration mechanism could rely on an intrinsic fitness for possible discoveries~ \cite{caldarelli2002scale, wang2013quantifying}, and one could add a parameter controlling for the strength of the coupling to further investigate the relation between the coupling of the two networks and the distribution of item popularity~\cite{gleeson2014simple}.

%
\medskip
\paragraph*{{\bf Conflict of Interest Statement}}
The authors declare that the research was conducted in the absence of any commercial or financial relationships that could be construed as a potential conflict of interest.

\paragraph*{{\bf Author Contributions}}
Conceived and designed the paper: II and VL. Performed the analysis and produced the figures: II. Wrote the paper: II and VL.

\paragraph*{{\bf Funding}}
I. I. acknowledges partial support from the UK RDRF - Urban Dynamics Lab under the EPSRC Grant No. EP/M023583/1 and from the ANR project DATAREDUX (ANR-19-CE46-0008). V. L. acknowledges support from the Leverhulme Trust Research Fellowship 278 “CREATE: the network components of creativity and success”.

\paragraph*{{\bf Acknowledgments}}
Figure icons made by Tana from www.flaticon.com.




%

\end{document}